\documentclass[a4paper,12pt]{article}

\usepackage{amssymb}
\usepackage{amsmath}
\usepackage{graphicx}
\usepackage[colorlinks=true,linkcolor=black,anchorcolor=black,citecolor=blue,filecolor=blue,menucolor=black,urlcolor=blue,breaklinks=true,pdfhighlight=/P,pdfmenubar=true,pdftoolbar=true,pdfpagelabels=true,pdfstartpage=1,pdfstartview=FitV,pdftitle={Counterflow in Evacuations},pdfsubject={Thema},pdfauthor={Kretz},pdfcreator={Kretz},pdfproducer={Kretz},pdfkeywords={pedestrian,crowd,simulation,traffic,VISSIM,VISWALK,counterflow,evacuation}]{hyperref}	
\usepackage[numbers,sort&compress]{natbib}
\usepackage{hypernat}				

\usepackage{simplemargins}
\setallmargins{1in}

\begin{document}

\title{Counterflow in Evacuations}
\author{Tobias Kretz\\
PTV Planung Transport Verkehr AG \\
Stumpfstr. 1\\
D-76131 Karlsruhe\\
\tt Tobias.Kretz@ptv.de}

\maketitle

\thispagestyle{empty}

\begin{abstract}
It is shown in this work that the average individual egress time and other performance indicators for egress of people from a building can be improved under certain circumstances if counterflow occurs. The circumstances include widely varying walking speeds and two differently far located exits with different capacity. The result is achieved both with a paper and pencil calculation as well as with a micro simulation of an example scenario. As the difficulty of exit signage with counterflow remains one cannot conclude from the result that an emergency evacuation procedure with counterflow would really be the better variant.

\end{abstract}

\section{Introduction and Description}

\begin{figure}[htbp]%
\begin{center}
\includegraphics[width=0.618\columnwidth]{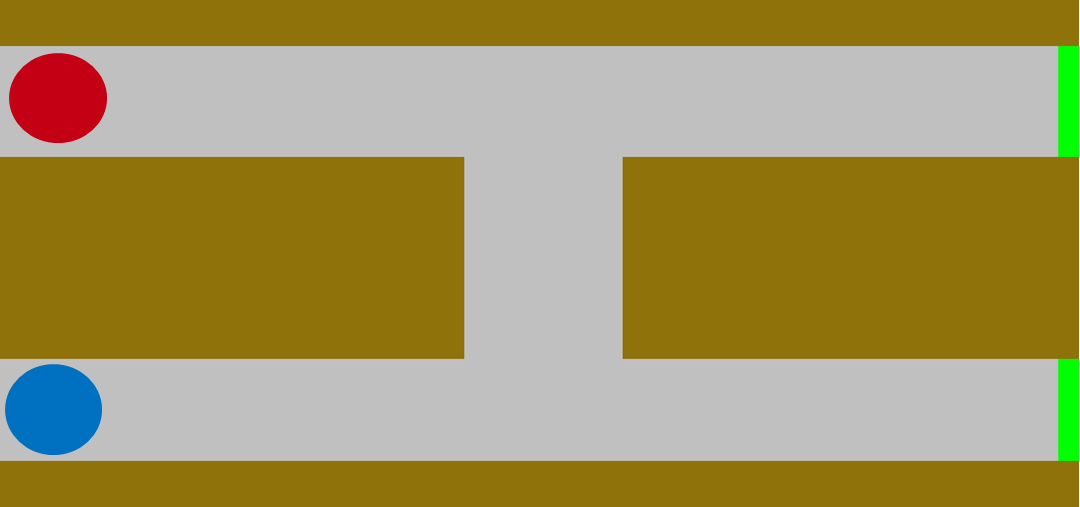}
\caption{Two groups of occupants (denoted by the blue and red spheres; walls are shown in brown) in a simple scenario with high symmetry: the exits (shown in green at the right side) are equally far away and have equal capacity. If the two groups have equal size in number and the occupants of both groups have comparable walking speeds it would not make sense to use the connecting corridor as it would imply a time delay.}%
\label{fig:simple}%
\end{center}
\end{figure}

It is usually assumed that during evacuations counterflow \cite{Simon1998,Muramatsu1999,Blue2000,Algadhi2001,Tajima2002,Isobe2004,Kretz2006d,Kretz2008d,Schadschneider2009,Schadschneider2009b} -- or bi-directional flows -- only occurs when occupants and rescue forces meet in opposite directions and that apart from that counterflows should not occur as they would inevitably imply that the time for evacuation is higher than necessary. This is surely true for a situation as shown in figure \ref{fig:simple}.

\begin{figure}[htbp]%
\begin{center}
\includegraphics[width=0.618\columnwidth]{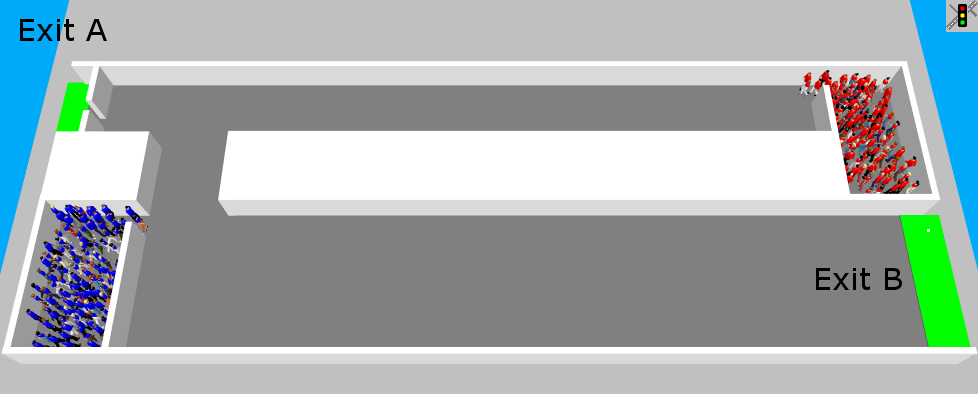} \\ \vspace{12pt}
\includegraphics[width=0.618\columnwidth]{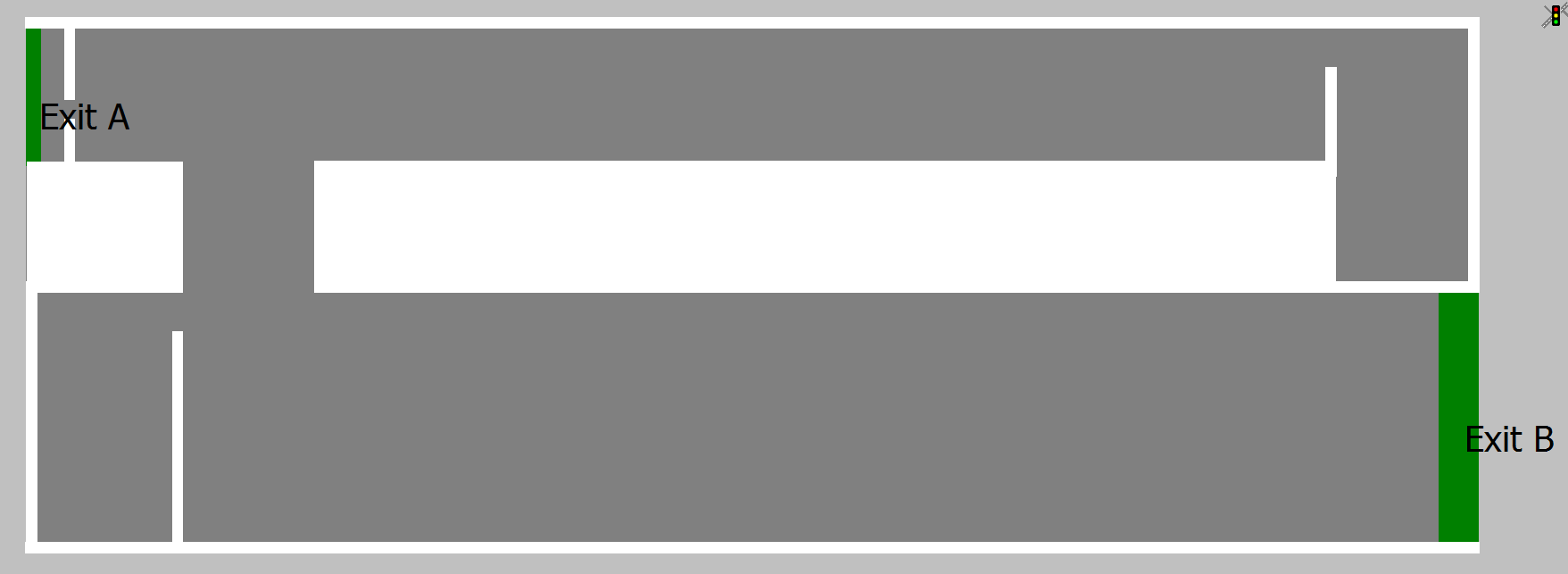}
\caption{3D and 2D view of the scenario: Two groups (blue and red) of occupants at the beginning of an evacuation process. The two available exits are marked in light green to the upper left and lower right. Note that here in front of exit A walls restrict the capacity.}%
\label{fig:scenario}%
\end{center}
\end{figure}

However, what if the exits have different capacities and are located at different walking distances from the two groups? What, furthermore, if the two groups have distinct walking speeds? Figure \ref{fig:scenario} shows such a situation. 

There are two groups of occupants shown in blue and red each in their own room. Each group has exactly 100 members. The two groups differ as the blue agents have a desired walking speed of $2.5\pm 0.2$ km/h (about 0.7 m/s) and the red agents $10.0 \pm 1.0$ km/h (about 2.8 m/s). Within the given boundaries the speeds are distributed equally among the population.

\begin{figure}[htbp]%
\begin{center}
\includegraphics[width=0.618\columnwidth]{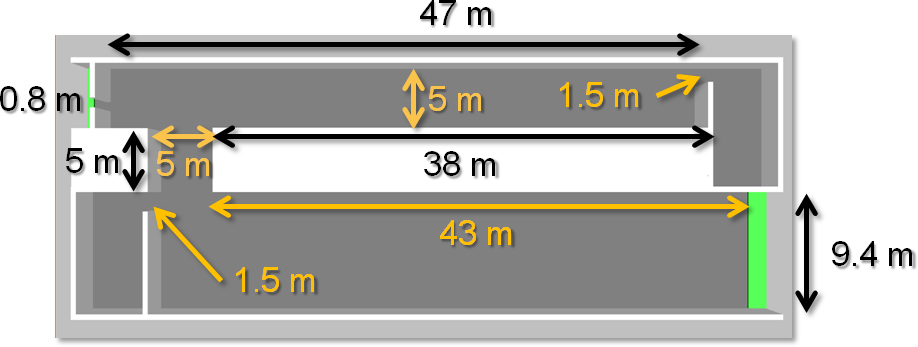} 
\caption{Measures}%
\label{fig:masse}%
\end{center}
\end{figure}

The two exits differ in two important aspects: the one on the upper left (exit A) is the closer one for both groups of occupants, but it has a restricted capacity -- the occupants need to pass a bottleneck before they reach the exit. The exit to the lower right (exit B) is more remote for both groups, but it has -- considering the total number of occupants -- nearly unlimited capacity. The geometric dimensions are shown in figure \ref{fig:masse}.

It is not difficult to see that the bottleneck in front of exit A will produce high total evacuation times as well as high average individual egress times if all occupants choose to use exit A. It therefore {\em might} make sense with regard to the user-equilibrium \cite{Bell1997} as well as the system optimum that a number of occupants use exit B. However, as exit B is rather far away, the {\em costs} to use it in terms of travel time are lower for the faster red agents. From this, one can hypothesize that to optimize at least some of the egress performance measures it makes sense to accept counterflow in the small corridor which connects the two main corridors.

Principally the distribution of pre-movement times as well as the individual choice of exits would produce a large number of variants. These have been reduced for this study to four by assuming a pre-movement time of zero and that all members of a group walk to the same exit.

This leaves four variants which are compared regarding the evacuation performance measure:
\begin{enumerate}
\item Shortest Path: blue group and red group both use exit A.
\item Maximum Capacity: blue group and red group both use exit B.
\item Separated: blue group uses exit B and red group uses exit A.
\item Counterflow: blue group uses exit A and red group uses exit B.
\end{enumerate}

This is shown as a matrix in table \ref{tab:properties}.

\begin{table}[htbp]%
\begin{center}
\begin{tabular}{cc|c|c} 
 $\ddots$ &     Blue &    Exit A     & Exit B           \\
Red       & $\ddots$ &               &                  \\ \hline
Exit A    &          & Shortest Path & Separated        \\ \hline
Exit B    &          & Counterflow   & Maximum Capacity \\
\end{tabular}
\caption{Main property of the egress strategy as a consequence of exit choice on the group level.}
\label{tab:properties}
\end{center}
\end{table}

\section{Paper and Pencil Calculation}
\begin{figure}[htbp]%
\begin{center}
\includegraphics[width=0.618\columnwidth]{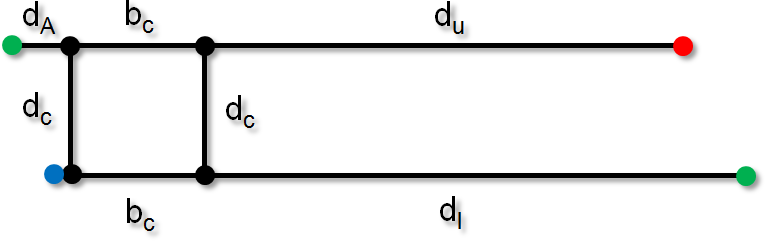} 
\caption{Simplified network display for the paper and pencil calculation. The lengths of the links are $d_A=4$ m, $b_c=5$ m, $d_u=38$ m, $d_c=5$ m, $d_l=43$ m.}%
\label{fig:network}%
\end{center}
\end{figure}

To investigate this hypothesis we first do a simple calculation of the total evacuation times for both groups, i.e. the time when the last pedestrian of each group has left the building. For this the geometry of figure \ref{fig:masse} is reduced to the network graph shown in figure \ref{fig:network}. In addition to the variables and their values given there, what is needed for the following calculations are the walking speed of a member of the red group $v_r = 2.78$ m/s, the walking speed of a member of the blue group $v_b = 0.69$ m/s, the assumed specific flow $j = 1.3$ 1/ms (in accordance with \cite{IMOMSC1238}), the width of exit A $b_A = 0.8$ m, the width of exit B: $b_B = 9.7$ m, and the number $N=100$ of the pedestrians in each of the groups. The widths of the doors of the rooms where the two groups start are equal: $b_r = b_b = 1.5$ m.

With these definitions and values the egress times for the two groups can be calculated. We begin with the simplest case.

\paragraph{Case 1: Separated}
In this case the bottleneck for the red group is exit A and the bottleneck for the blue group is the door of their starting room. Therefore the evacuation times result as
\begin{eqnarray}
T_{red}&=&\frac{d_u + b_c + d_A}{v_r} + \frac{N}{j b_A} = 113.1 \text{ s}\\
T_{blue}&=&\frac{N}{j b_b} + \frac{b_c + d_l}{v_b} = 120.8 \text{ s}
\end{eqnarray}

\paragraph{Case 2: Maximum Capacity}

We neglect that the members of the red group might slow down when walking around the corners and when they have to overtake the members of the blue group. For the blue group as well we neglect any negative impact from the fact that they now have to share the wide (lower) corridor with the red group. Therefore their evacuation time is calculated in the same way as in the previous case.
\begin{eqnarray}
T_{red}&=&\frac{N}{j b_r} + \frac{d_u + d_c + d_l}{v_r} = 82.2 \text{ s}\\
T_{blue}&=&\frac{N}{j b_b} + \frac{b_c + d_l}{v_b} = 120.8 \text{ s}
\end{eqnarray}

\paragraph{Case 3: Shortest Path}
Here both groups queue before the narrow exit A. The evacuation times depend much on the assumption which group moves first through the bottleneck. If it was the red group, for the red group an evacuation time as in case 1 would result. If it was the blue group an evacuation time as in case 4 would follow. For the total evacuation time the worst case assumption is that both groups move alternating. This maximizes the time for both groups and gives -- in this type of calculation -- equal results.
\begin{eqnarray}
T_{red}&=&\text{Min}\left(\frac{d_u + b_c + d_A}{v_r},\frac{d_c + d_A}{v_b}\right) + \frac{2N}{j b_A} = 205.3 \text{ s}\\
T_{blue}&=&\text{Min}\left(\frac{d_u + b_c + d_A}{v_r},\frac{d_c + d_A}{v_b}\right) + \frac{2N}{j b_A} = 205.3 \text{ s}
\end{eqnarray}

\paragraph{Case 4: Counterflow}
The puzzling question in this case is with what factor one should take care for the fact of counterflow. For this simple calculation here we decide to do not at all so, i.e. assume that the counterflow is as efficient as if it was uni-directional. First the connecting corridor is relatively wide compared to the two doors of the starting rooms, second, from this simple calculation it does not follow how long the counterflow situation persists, and third with an assumed optimal flow from the result it can become apparent how large the loss in flow efficiency could be, before the conclusions of the calculation would have to be modified.

\begin{eqnarray}
T_{red}&=&\frac{N}{j b_r} + \frac{d_u + d_c + d_l}{v_r} = 82.2 \text{ s}\\
T_{blue}&=&\frac{d_c + d_A}{v_b} + \frac{N}{j b_A} = 109.2 \text{ s}
\end{eqnarray}

Table \ref{tab:resultspap} summarizes these results. It shows that the counterflow strategy performs best.

\begin{table}[htbp]%
\begin{center} 
\begin{tabular}{cc|cc|cc} 
 $\ddots$      &Blue&             &Exit A       &            &Exit B         \\
Red   &$\ddots$&             &              &            &               \\ \hline
      &        &             &205.3 &            &120.8  \\
Exit A&        &205.3 &              &113.1&              \\ \hline
      &        &             & 109.2&            &120.8 \\
Exit B&        & 82.2&              &82.2&               \\
\end{tabular}
\caption{Summary of results of paper and pencil calculation}
\label{tab:resultspap}
\end{center}
\end{table}

\section{Simulation}

Motivated by this result but still careful for having done many simplifications, we now simulate the scenario using VISWALK (the pedestrian module of VISSIM) \cite{Kretz2008b,Fellendorf2010,VISSIM2010}. The pedestrian dynamics model of VISWALK is the variant of the Social Force Model introduced as {\em elliptical variant II} in \cite{Johansson2007,Helbing2009} with minor modifications and extensions for specific situations.

At the beginning of the simulation the occupants are set into the simulation equally distributed over their room. 

Each of the four variants has been simulated 20 times with different random numbers for the stochastic term of the Social Force Model. Table \ref{tab:results1} shows the average (over simulation runs) of averages of individual egress times.

\begin{figure}[htbp]%
\begin{center}
\includegraphics[width=0.45\columnwidth]{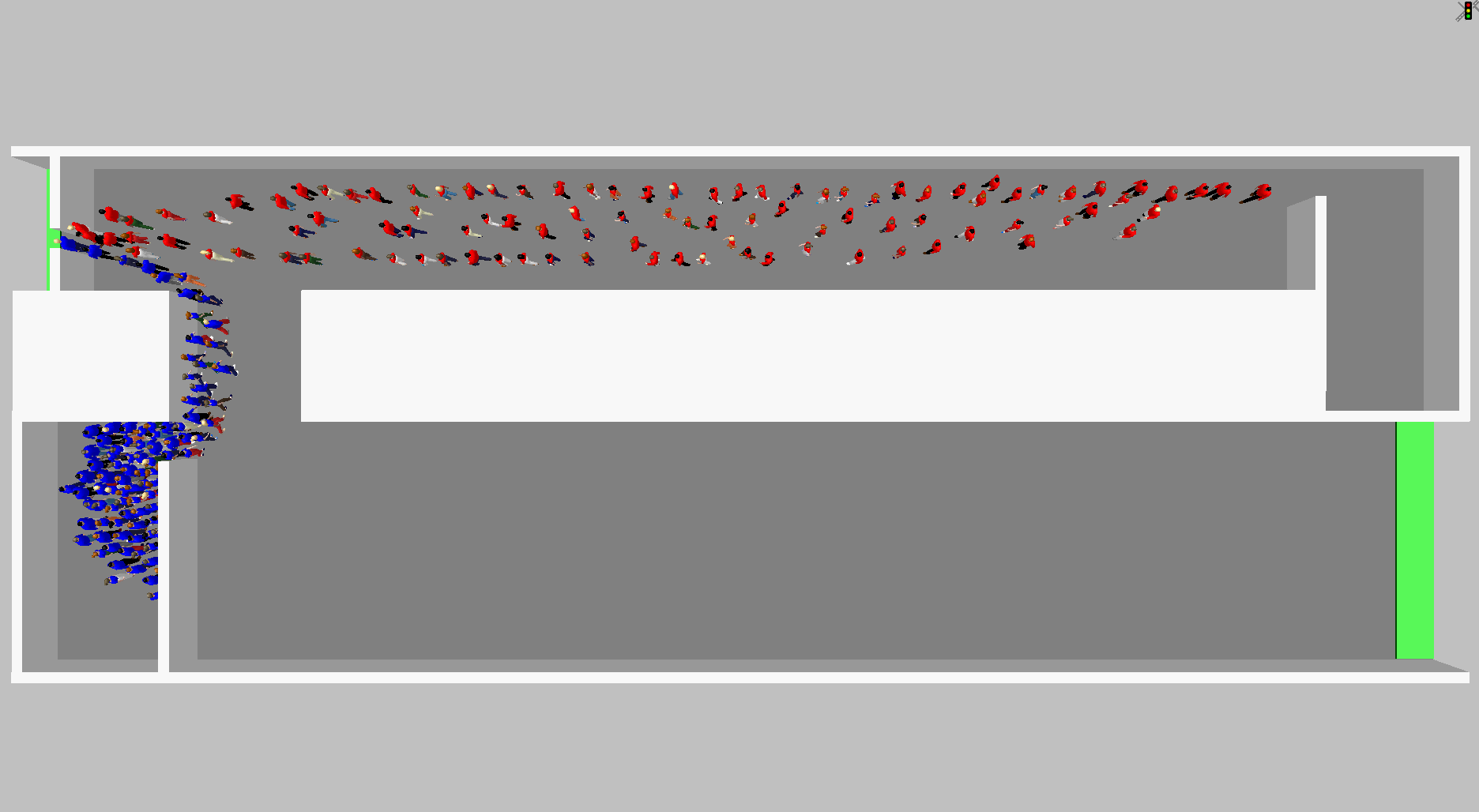} \hspace{12pt}
\includegraphics[width=0.45\columnwidth]{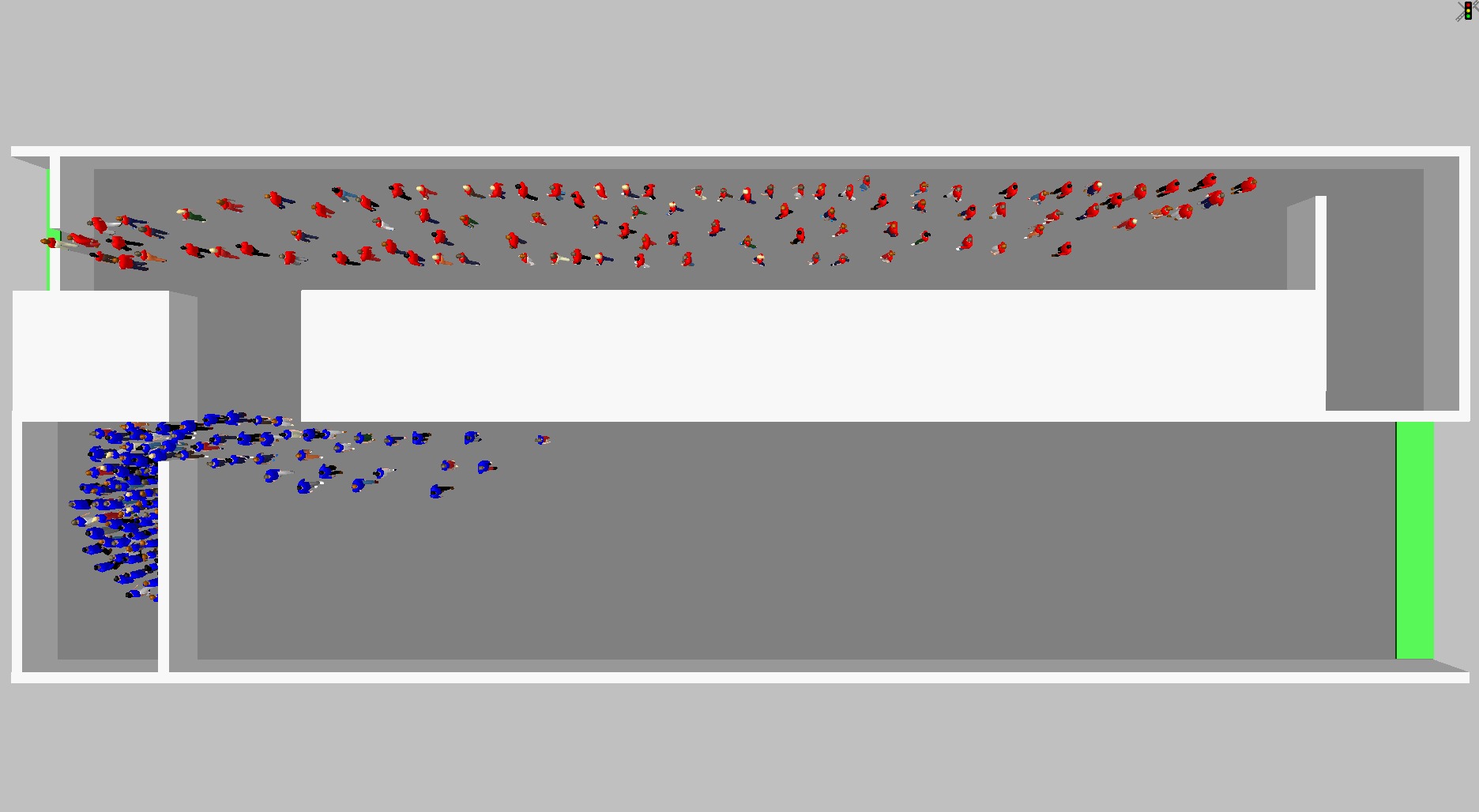} \\ \vspace{12pt}
\includegraphics[width=0.45\columnwidth]{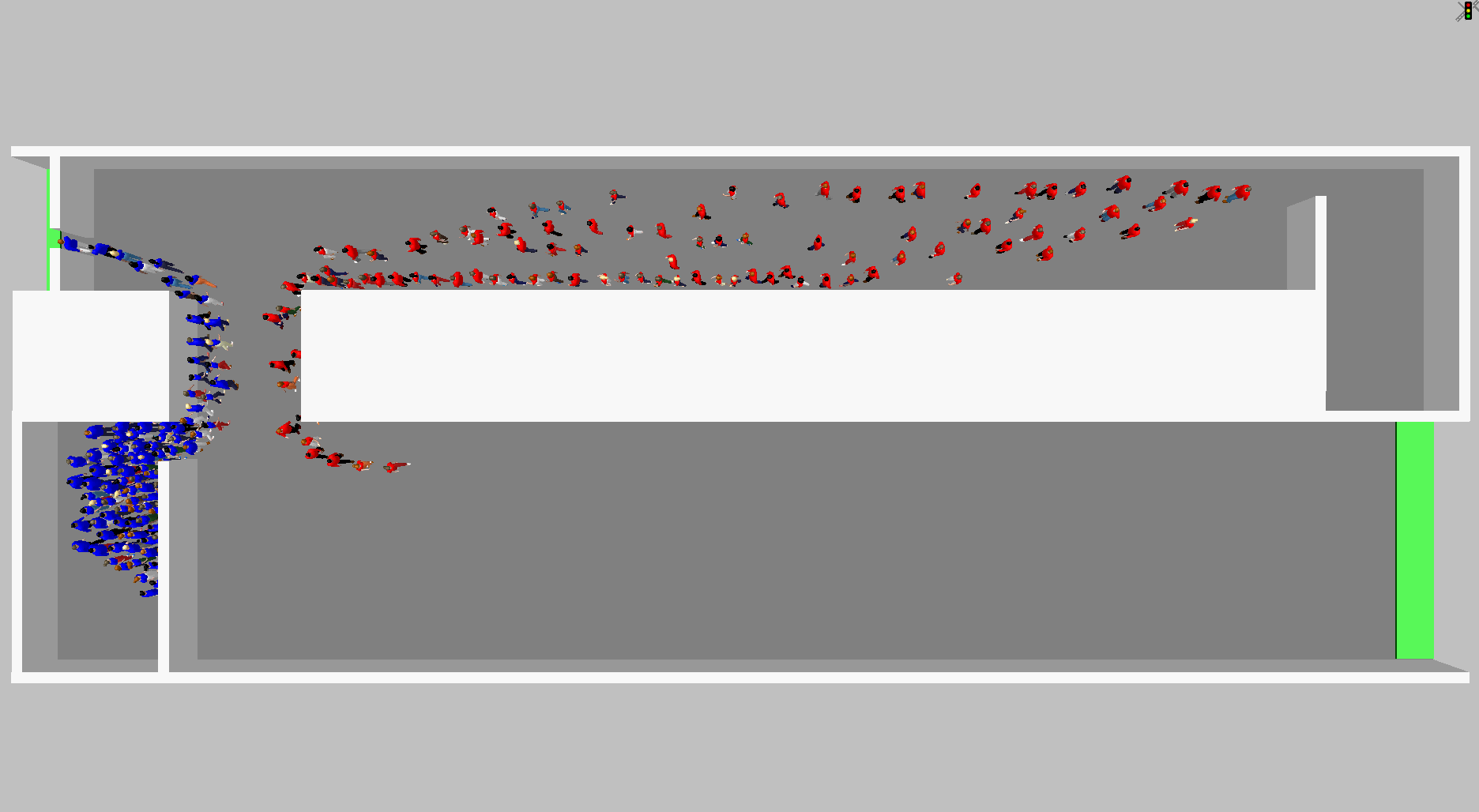} \hspace{12pt}
\includegraphics[width=0.45\columnwidth]{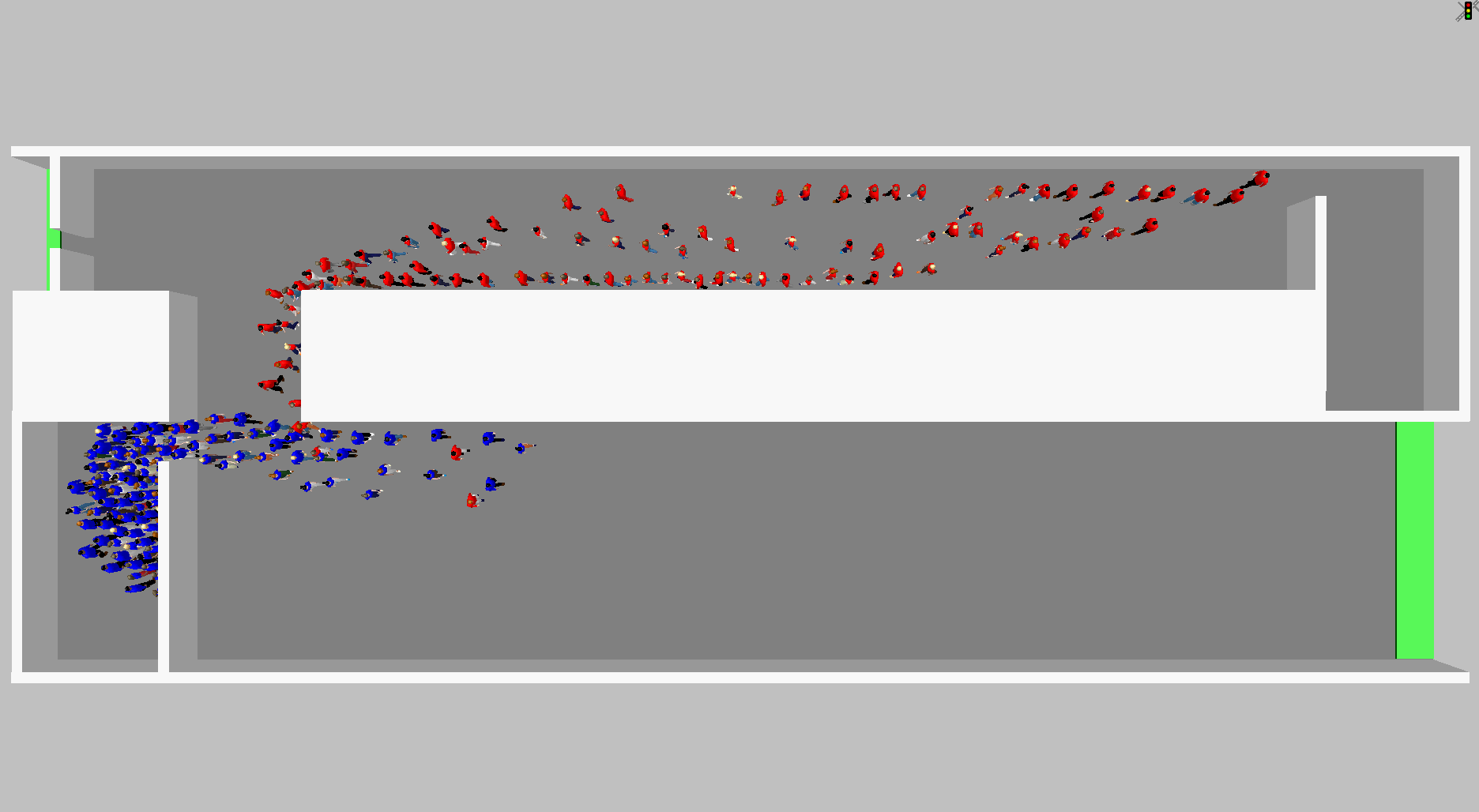} 
\caption{Situation after 20 seconds for all four variants (the four figures are placed according to the matrix of table \ref{tab:properties}).}%
\label{fig:20sec}%
\end{center}
\end{figure}

\begin{figure}[htbp]%
\begin{center}
\includegraphics[width=0.45\columnwidth]{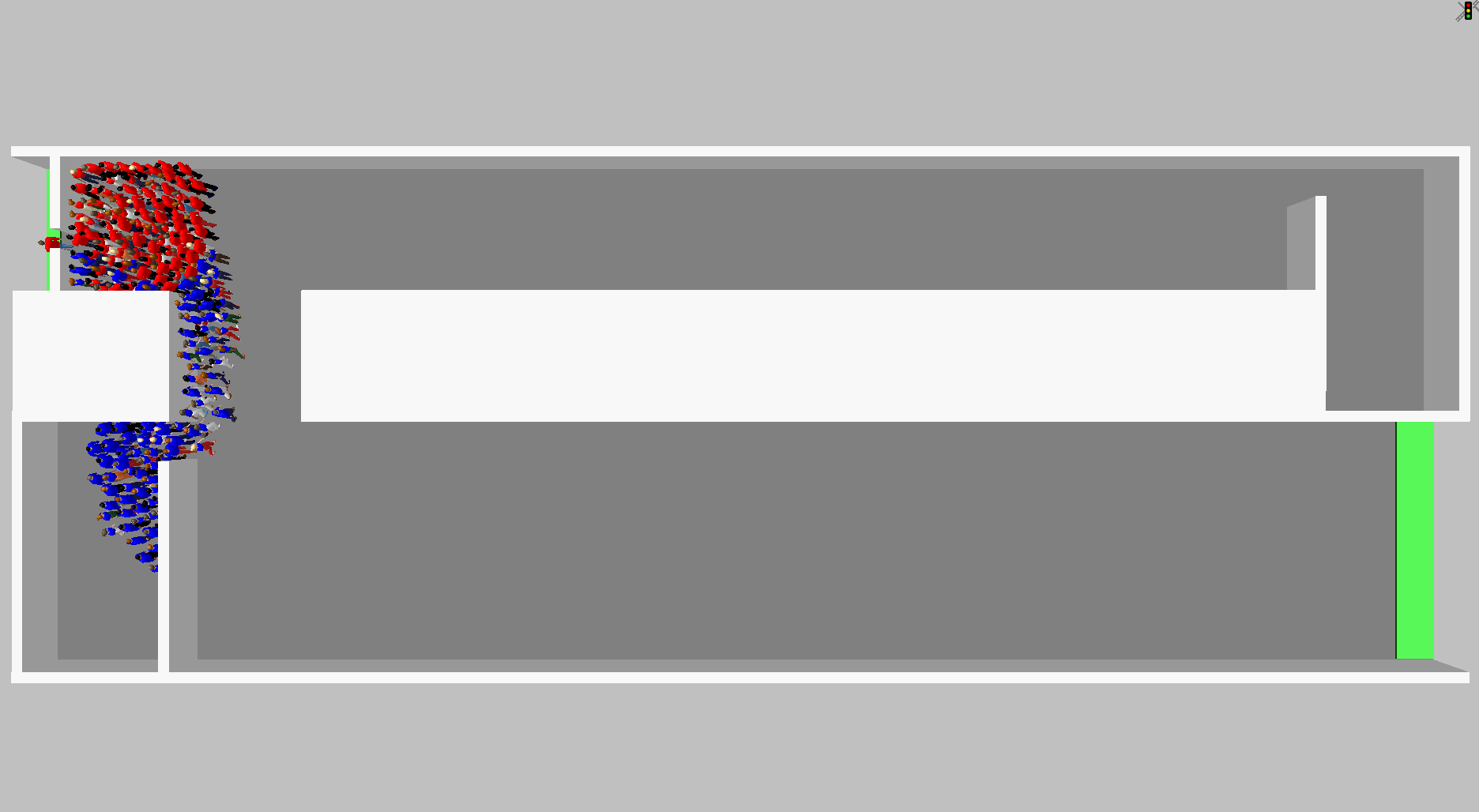} \hspace{12pt}
\includegraphics[width=0.45\columnwidth]{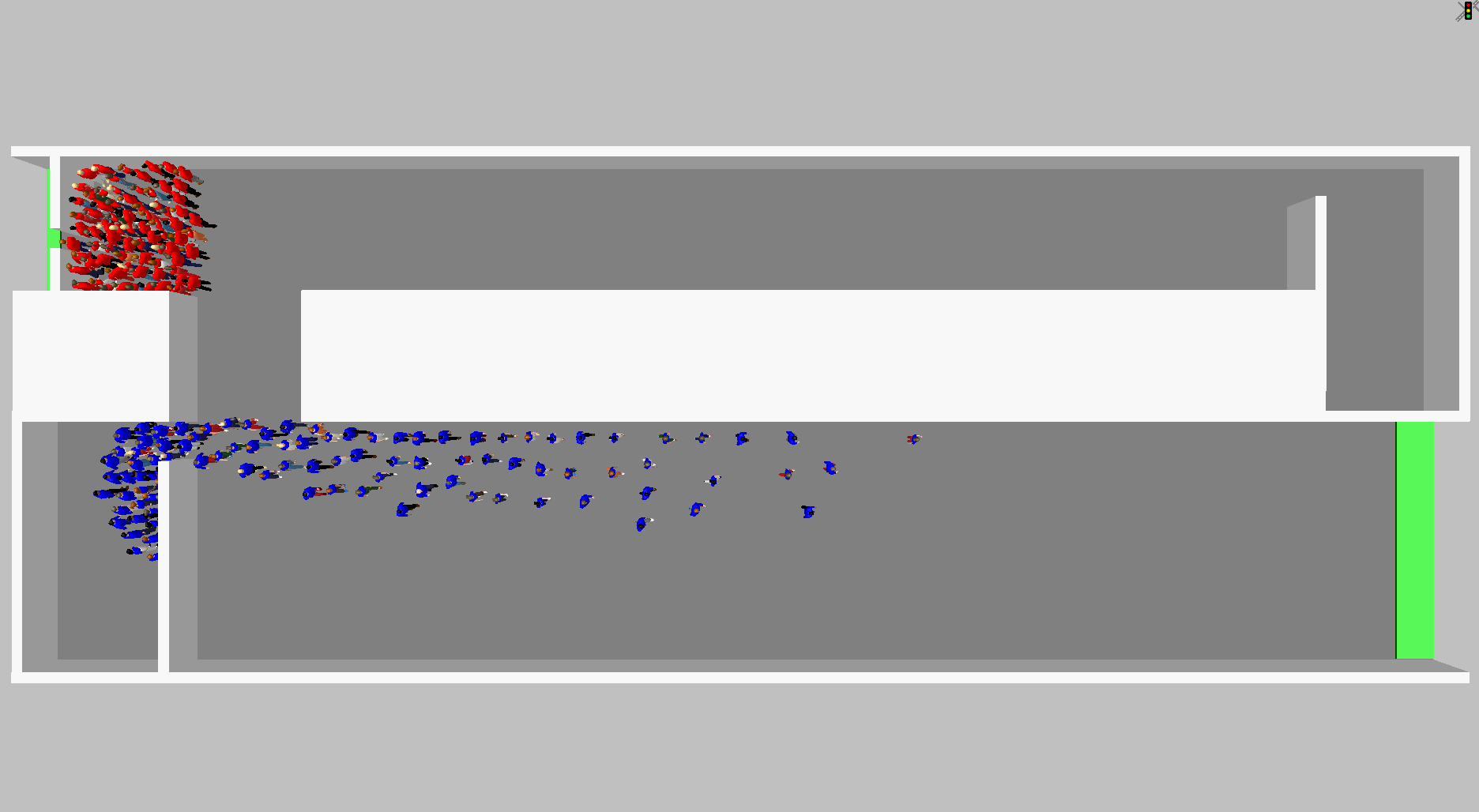} \\ \vspace{12pt}
\includegraphics[width=0.45\columnwidth]{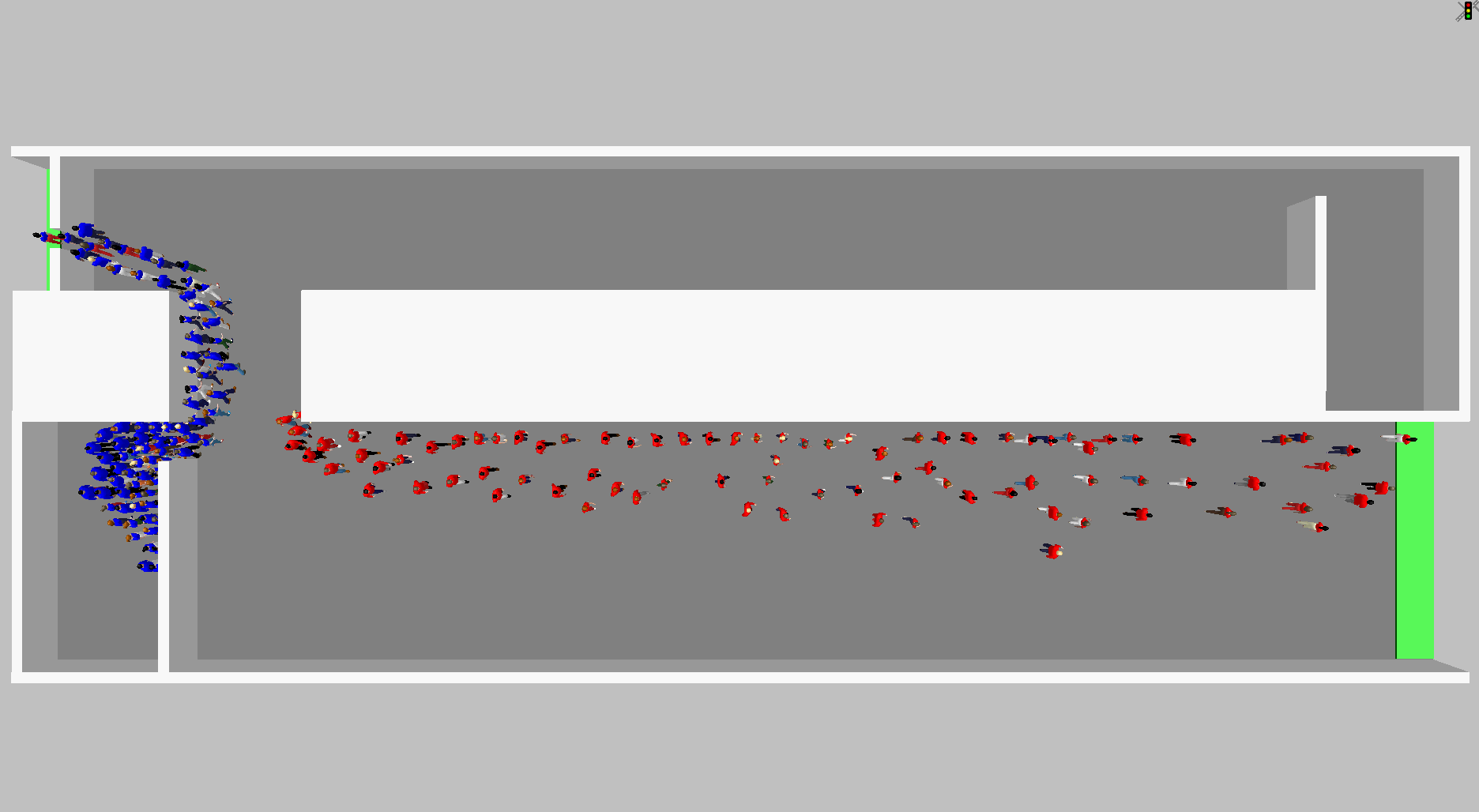} \hspace{12pt}
\includegraphics[width=0.45\columnwidth]{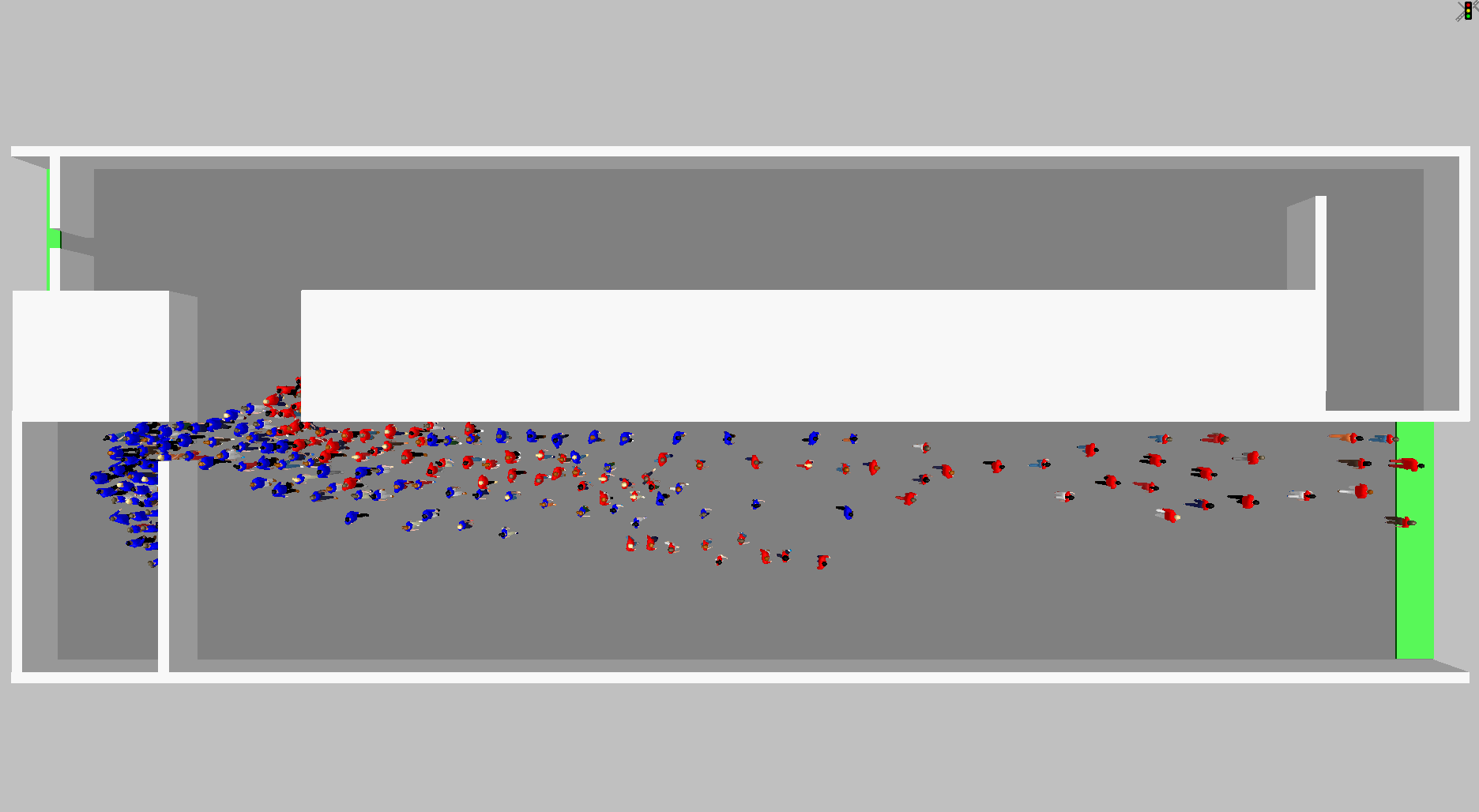} 
\caption{Situation after 40 seconds for all four variants (the four figures are placed according to the matrix of table \ref{tab:properties}).}%
\label{fig:40sec}%
\end{center}
\end{figure}

\begin{figure}[htbp]%
\begin{center}
\includegraphics[width=0.45\columnwidth]{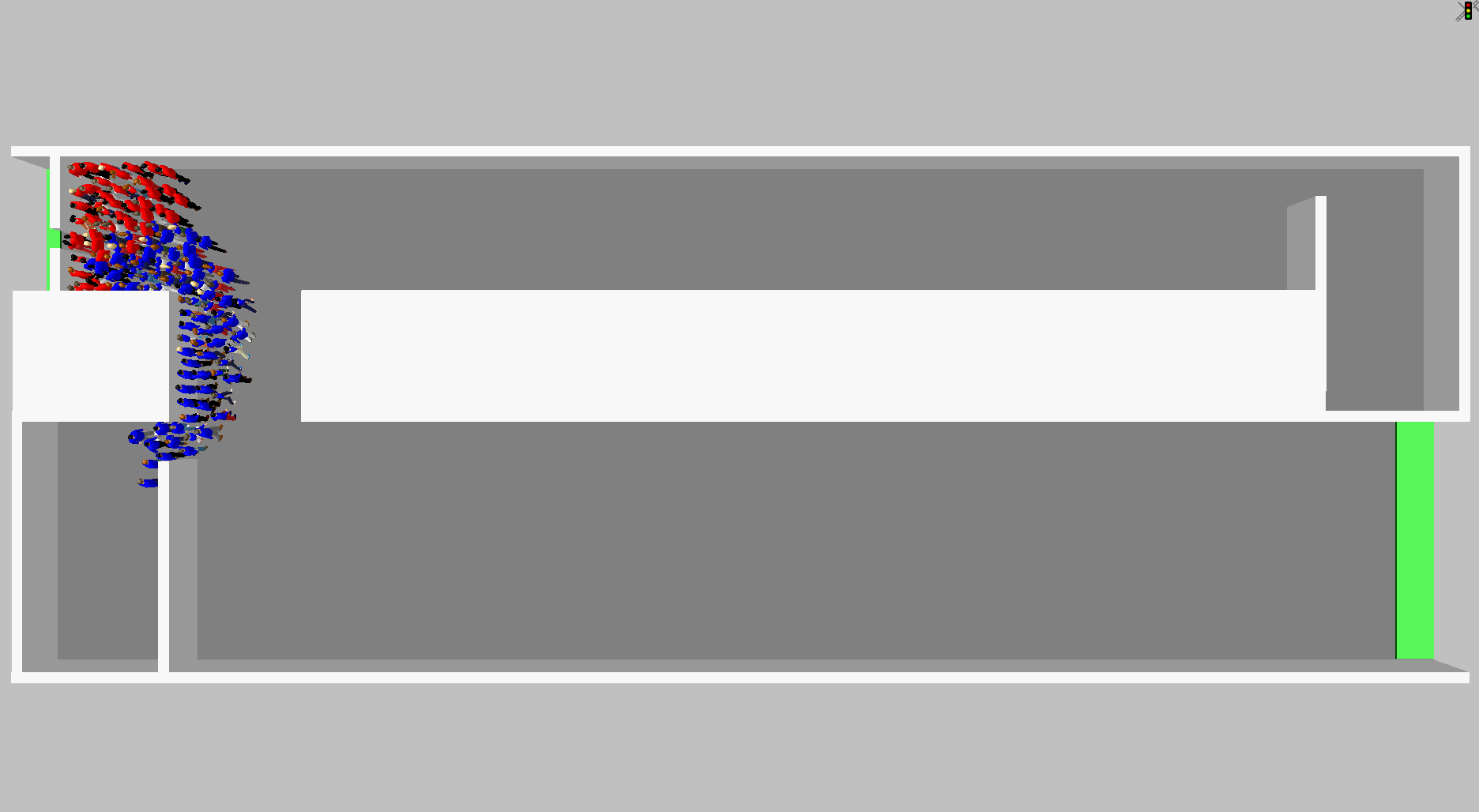} \hspace{12pt}
\includegraphics[width=0.45\columnwidth]{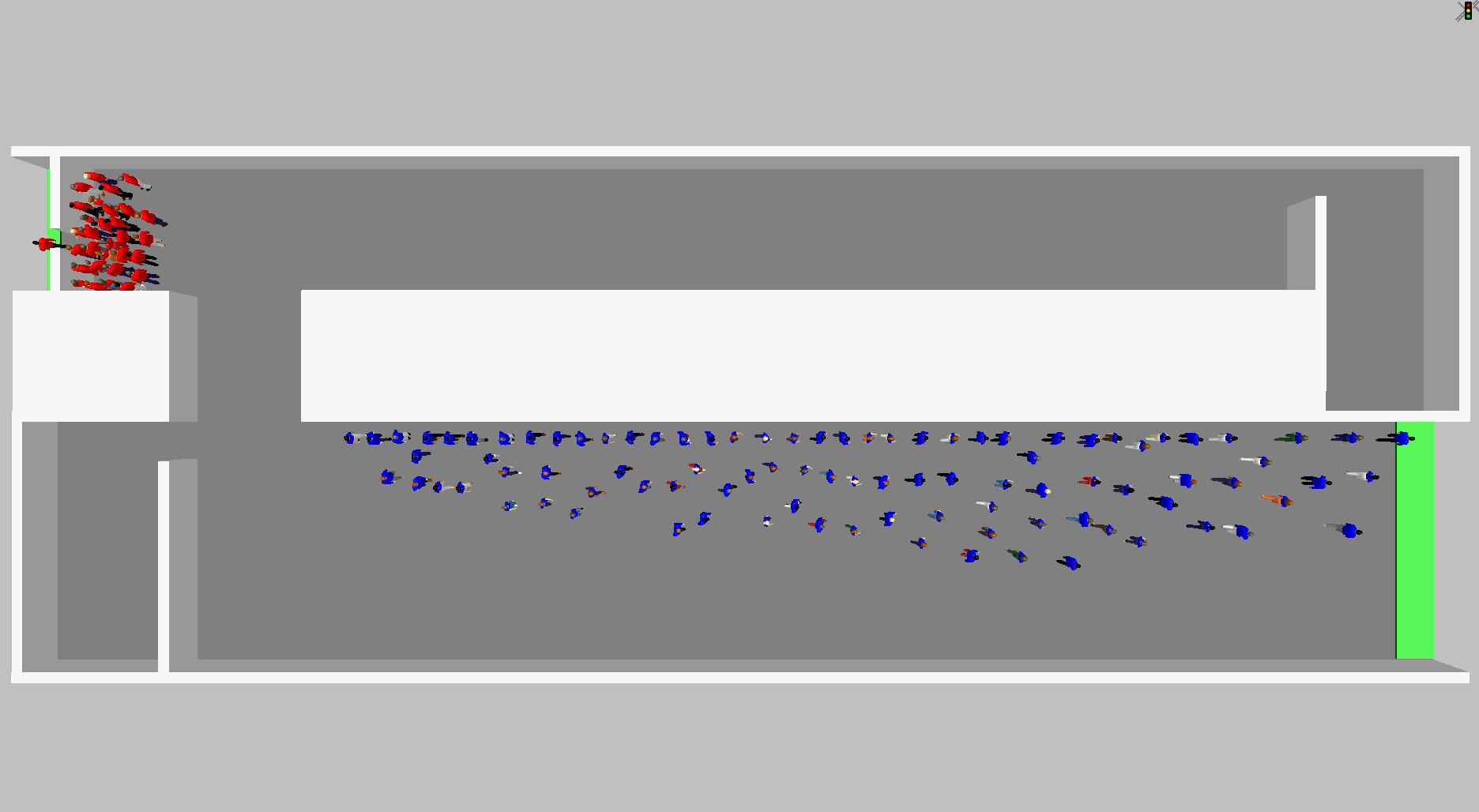} \\ \vspace{12pt}
\includegraphics[width=0.45\columnwidth]{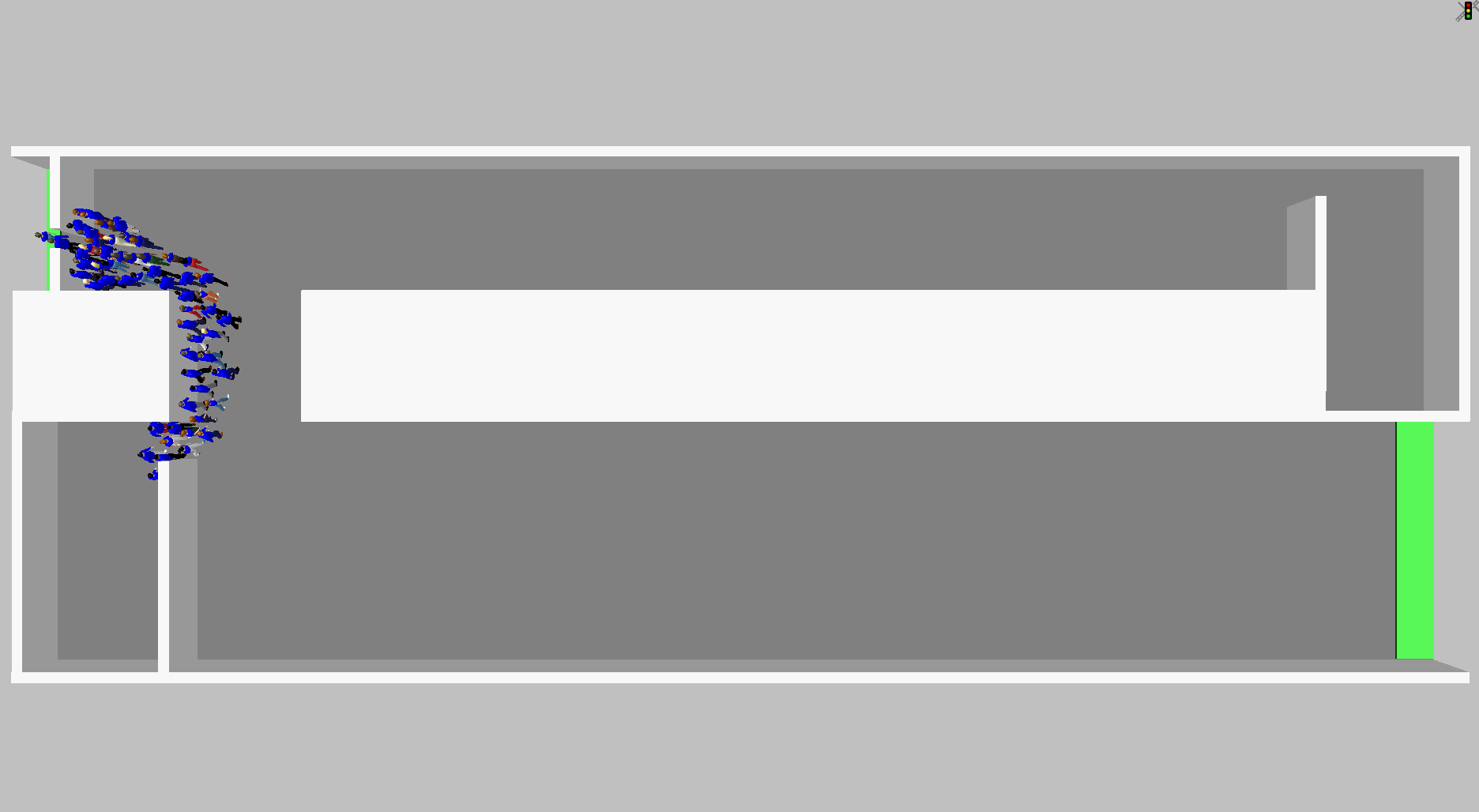} \hspace{12pt}
\includegraphics[width=0.45\columnwidth]{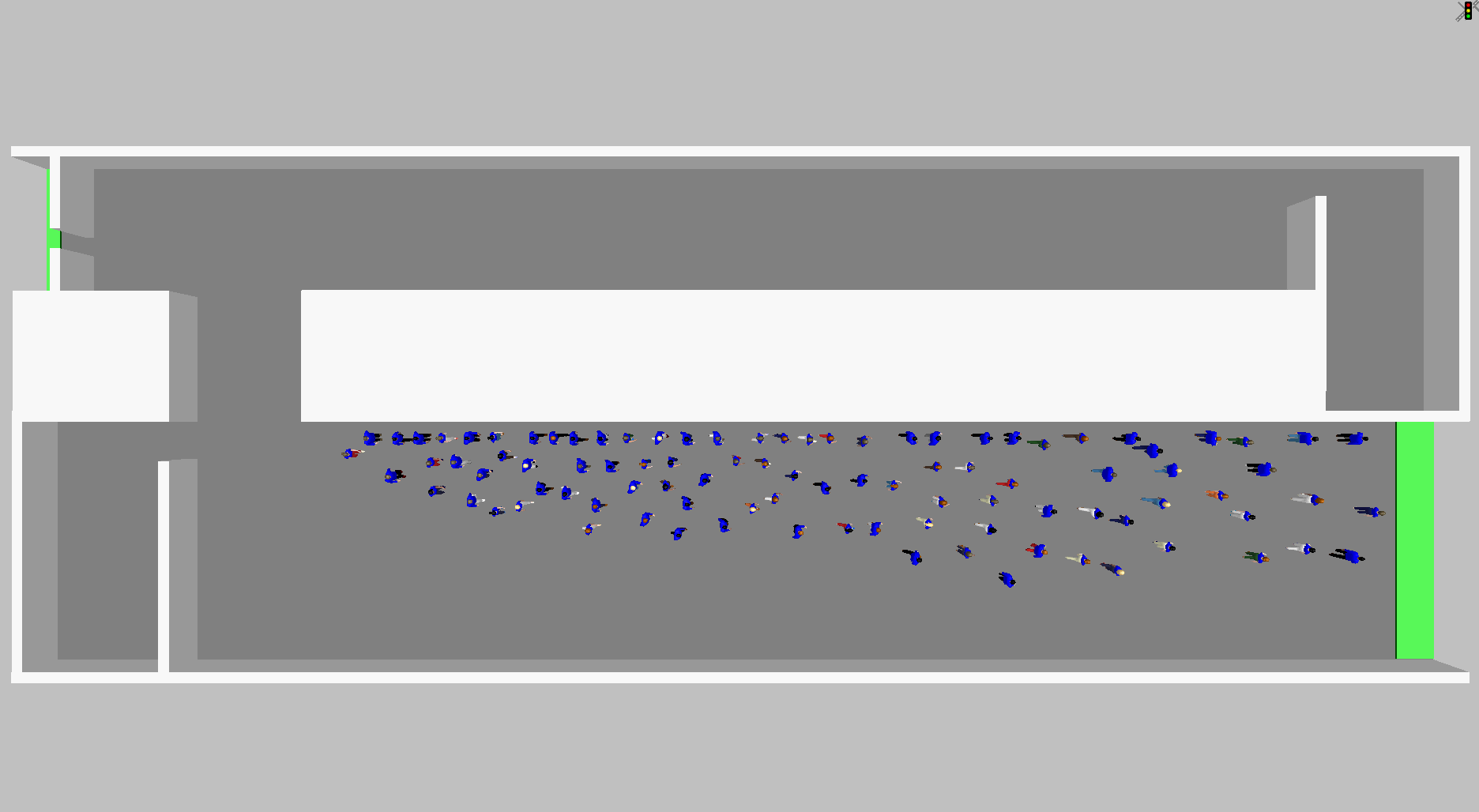} 
\caption{Situation after 80 seconds for all four variants (the four figures are placed according to the matrix of table \ref{tab:properties}).}%
\label{fig:80sec}%
\end{center}
\end{figure}

\begin{figure}[htbp]%
\begin{center}
\includegraphics[width=0.45\columnwidth]{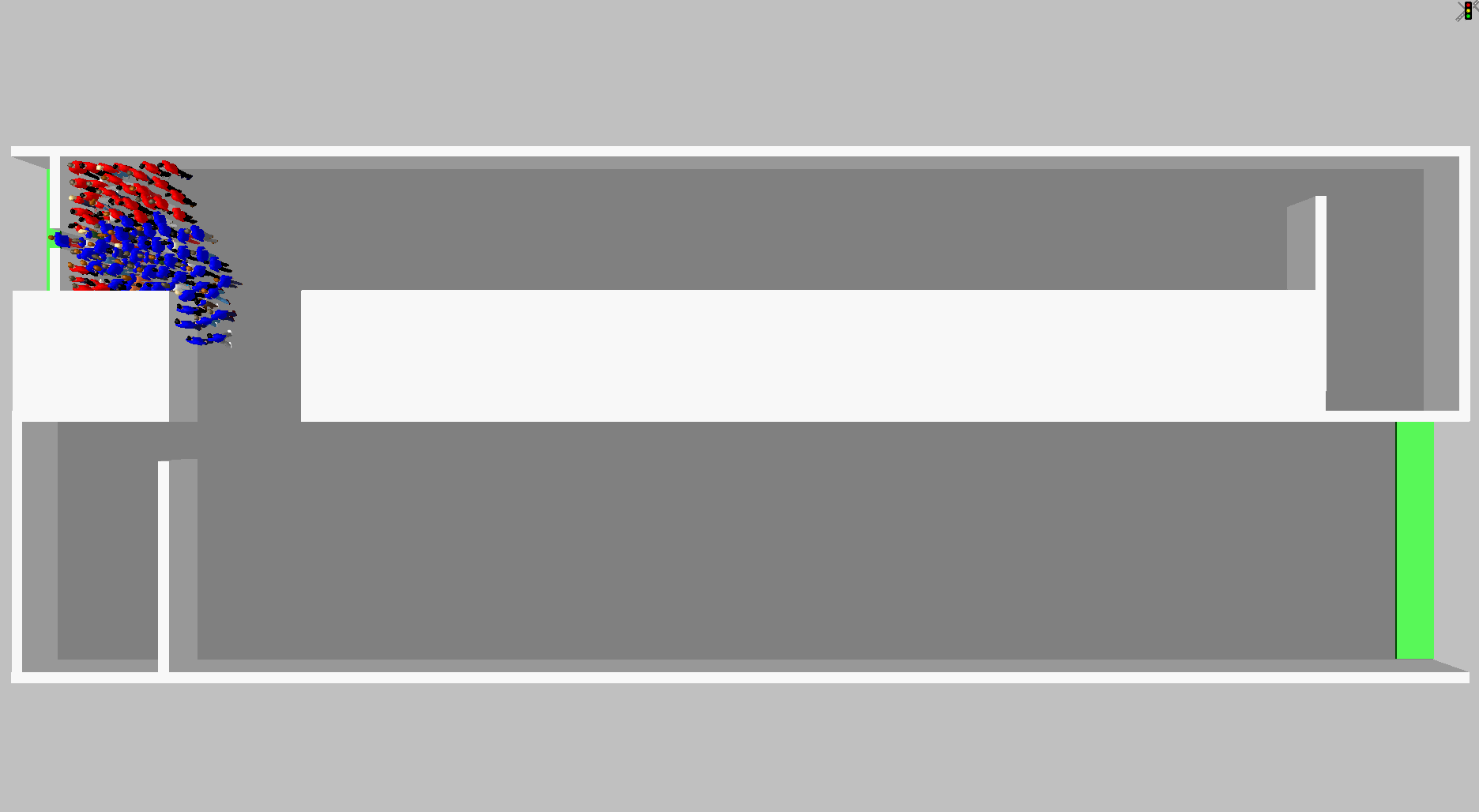} \hspace{12pt}
\includegraphics[width=0.45\columnwidth]{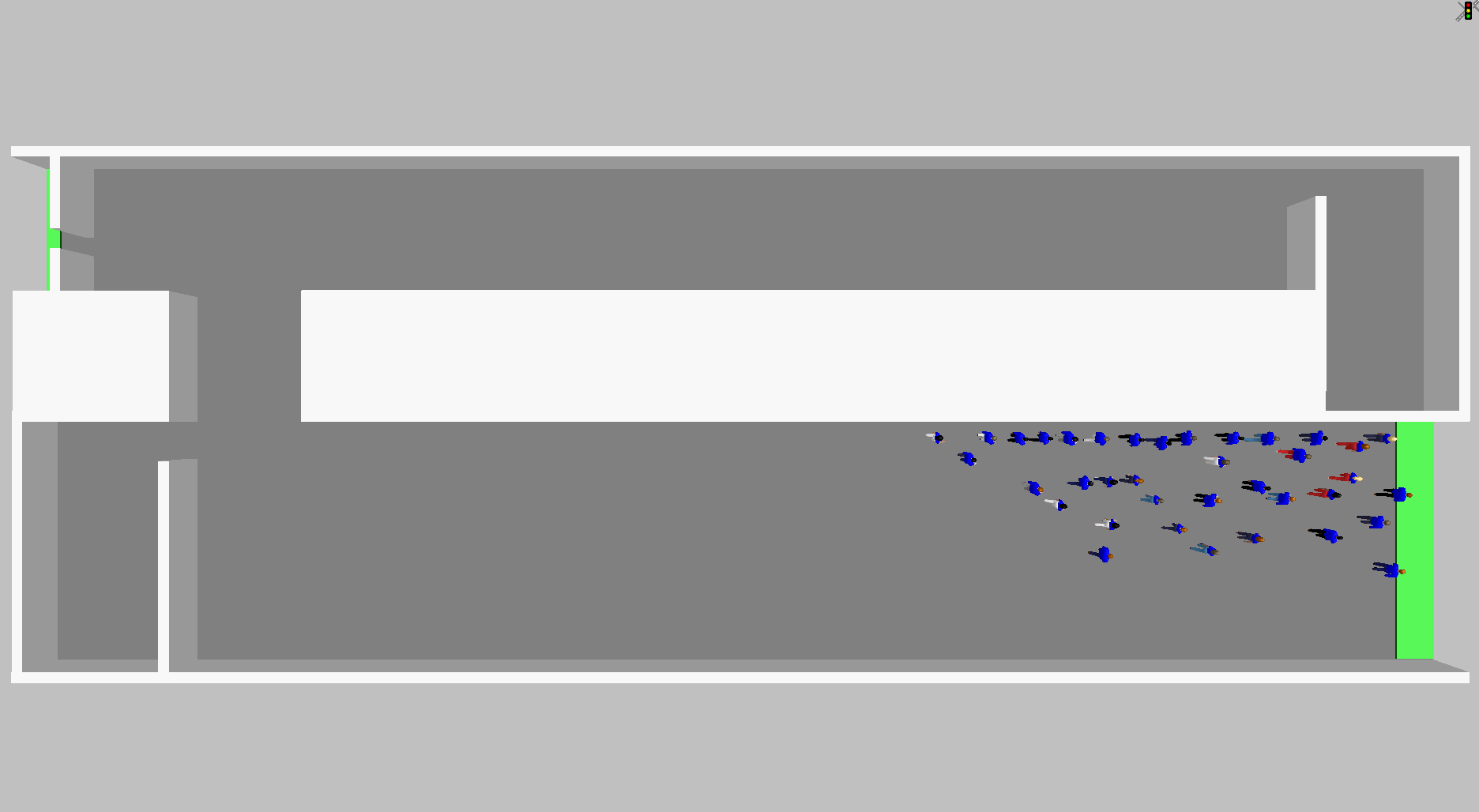} \\ \vspace{12pt}
\includegraphics[width=0.45\columnwidth]{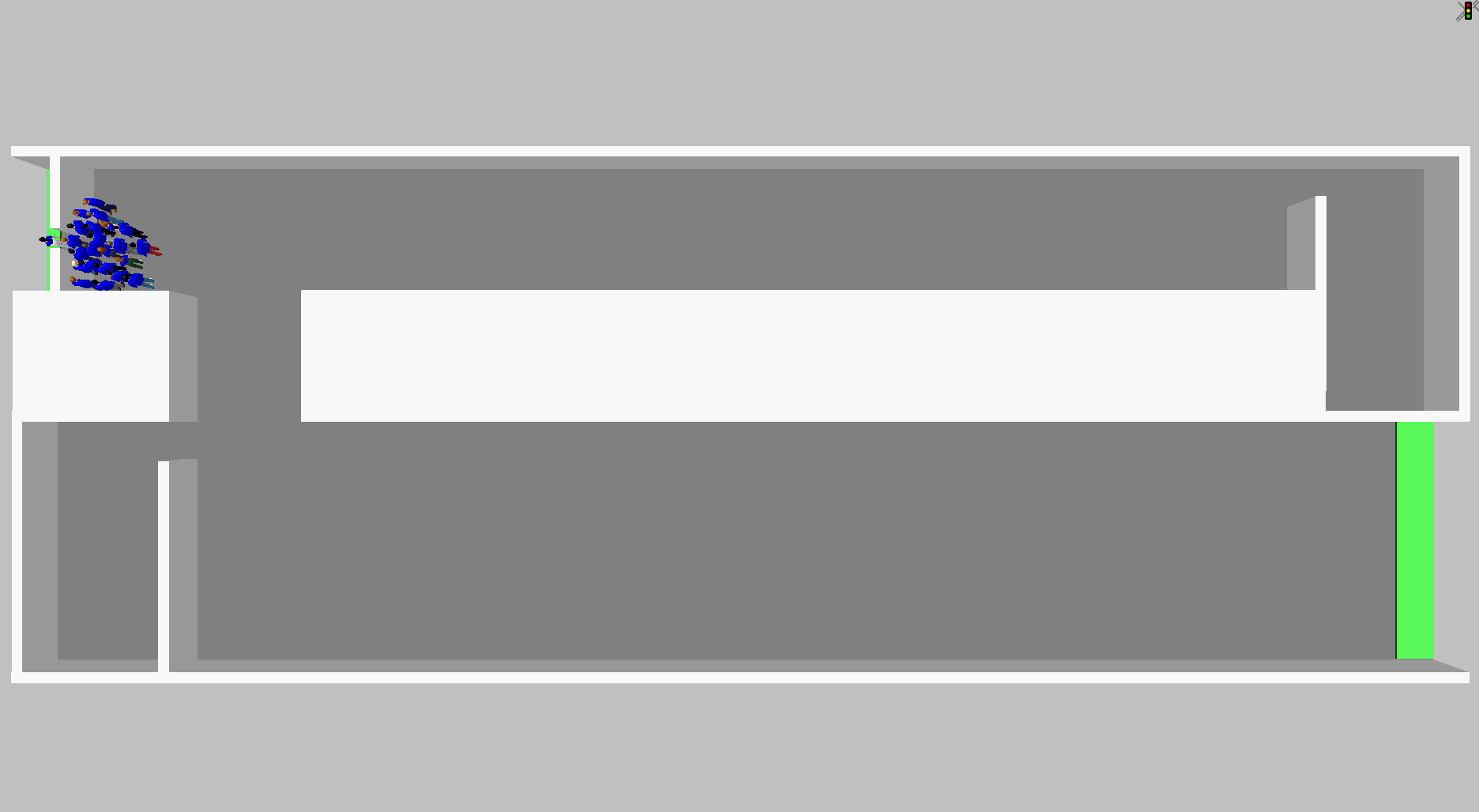} \hspace{12pt}
\includegraphics[width=0.45\columnwidth]{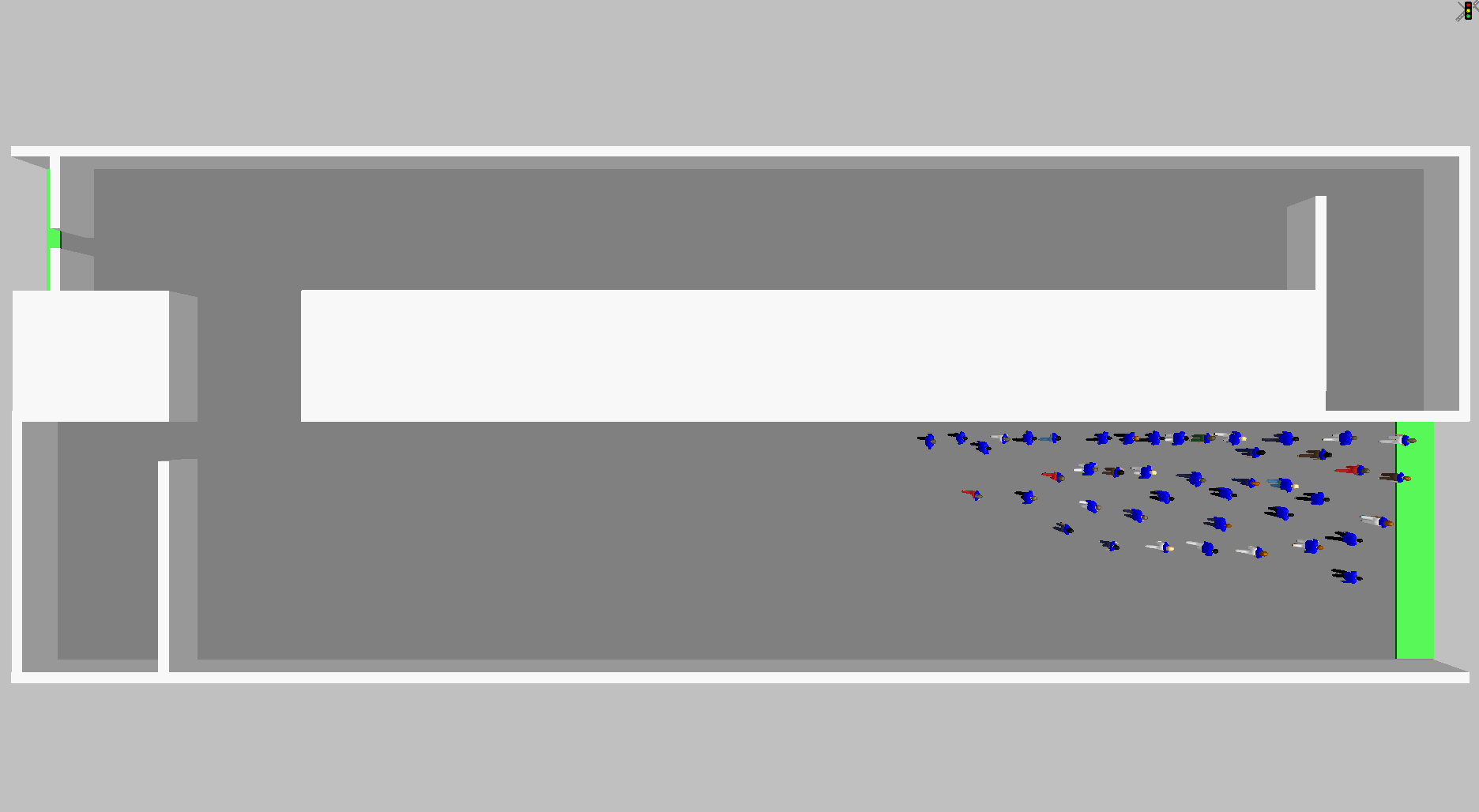} 
\caption{Situation after 120 seconds for all four variants (the four figures are placed according to the matrix of table \ref{tab:properties}).}%
\label{fig:120sec}%
\end{center}
\end{figure}

Figures \ref{fig:20sec} to \ref{fig:120sec} show screen captures of the animated simulation runs. From the 20 simulation runs that were carried out those four were selected (respectively their random number generator seed value), which were closest to the average with respect to the average individual egress time for both the blue as well as the red agents.

\begin{table}[htbp]%
\begin{center}
\begin{tabular}{cc|cc|cc} 
 $\ddots$      &Blue&             &Exit A       &            &Exit B         \\
Red   &$\ddots$&             &             &            &               \\ \hline
      &        &             &124.5$\pm$3.1&            &112.6$\pm$1.1  \\
Exit A&        &104.9$\pm$3.8&             &62.7$\pm$2.2&               \\ \hline
      &        &             & 87.1$\pm$1.8&            &114.7$\pm$1.1  \\
Exit B&        & 47.9$\pm$0.5&             &49.8$\pm$0.7&               \\
\end{tabular}
\caption{Average of averages of individual egress times in seconds.}
\label{tab:results1}
\end{center}
\end{table}

These results show that for the blue as well as for the red group the case with counterflow leads to the smallest value of average individual egress times, with regard to this measure this therefore obviously is the system optimum. It can easily be seen that with table \ref{tab:results1} as cost matrix in an iterated game \cite{Kreps1992,Camerer2003,Osborne2004,Kretz2011d} the case with counterflow as well emerges as stable user equilibrium.

\begin{table}[htbp]%
\begin{center} 
\begin{tabular}{cc|cc|cc} 
 $\ddots$      &Blue&             &Exit A       &            &Exit B         \\
Red   &$\ddots$&             &              &            &               \\ \hline
      &        &             &189.3$\pm$4.6 &            &143.5$\pm$1.6  \\
Exit A&        &221.2$\pm$4.9 &              &107.5$\pm$4.2&              \\ \hline
      &        &             & 146.5$\pm$2.8&            &145.4$\pm$1.6 \\
Exit B&        & 58.1$\pm$0.6&              &61.3$\pm$1.2&               \\
\end{tabular}
\caption{Average of simulation times when 95\% of all agents had reached one of the two destination areas.}
\label{tab:results95}
\end{center}
\end{table}

With respect to the time when 95\% of the agents (i.e. 190 agents) have reached one of the two destination areas for the red group again the counterflow strategy would be clearly the best one, while within the variance of results for the blue group ``counterflow'', ``separated'', and ``maximum capacity'' are equally well suited, only ``shortest path'' is worse. If one neglects the statistical scatter of results and only considers the mean value then the ``Maximum Capacity'' strategy would be the user equilibrium. 

\begin{table}[htbp]%
\begin{center} 
\begin{tabular}{cc|cc|cc} 
 $\ddots$      &Blue&             &Exit A       &            &Exit B         \\
Red   &$\ddots$&             &              &            &               \\ \hline
      &        &             &206.1$\pm$6.3 &            &150.1$\pm$2.1  \\
Exit A&        &229.5$\pm$4.9 &              &116.1$\pm$4.6&              \\ \hline
      &        &             & 158.0$\pm$3.4&            &151.9$\pm$1.9 \\
Exit B&        & 59.8$\pm$0.6&              &63.5$\pm$1.3&               \\
\end{tabular}
\caption{Average of total evacuation times (all agents have left the simulation).}
\label{tab:results100}
\end{center}
\end{table}

Finally for the case of total evacuation times as shown in table \ref{tab:results100} -- only considering mean values and neglecting the standard deviations -- the counterflow strategy remains the best one for the blue group, the sum of times is best for the maximum capacity strategy which is also the user equilibrium strategy. For the blue group the separated strategy scores best. It is not surprising that with increasing number of pedestrians within the same geometry the optimum choice shifts toward a maximum capacity strategy.

\begin{figure}[htbp]%
\begin{center}
\includegraphics[width=0.618\columnwidth]{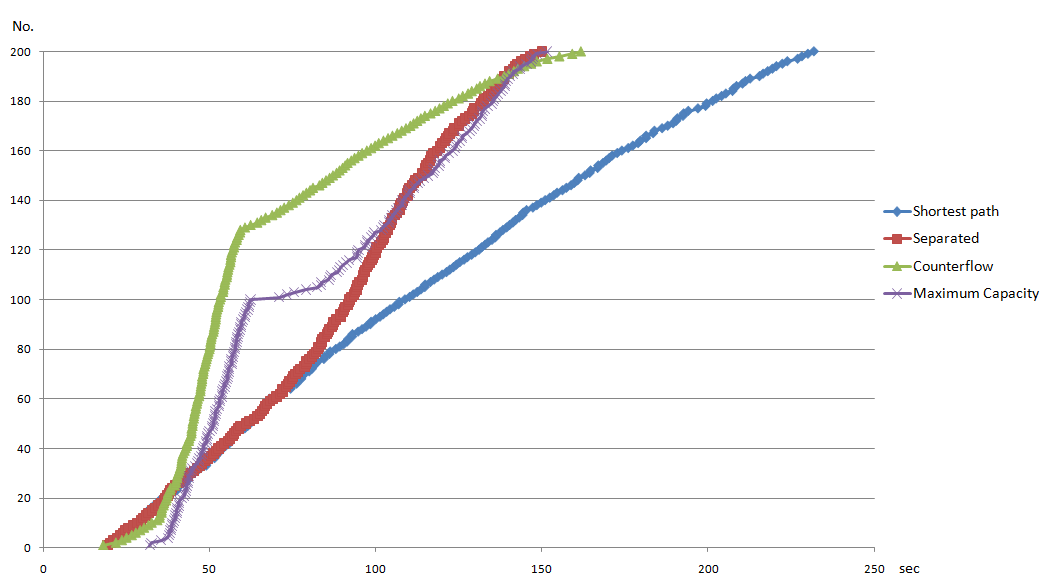}
\caption{Number of agents that have arrived at their destination in dependence of the simulation time (evacuation graph). This plot shows most clearly that for a wide range the counterflow strategy is the best one.}%
\label{fig:evacboth}%
\end{center}
\end{figure}

\begin{figure}[htbp]%
\begin{center}
\includegraphics[width=0.618\columnwidth]{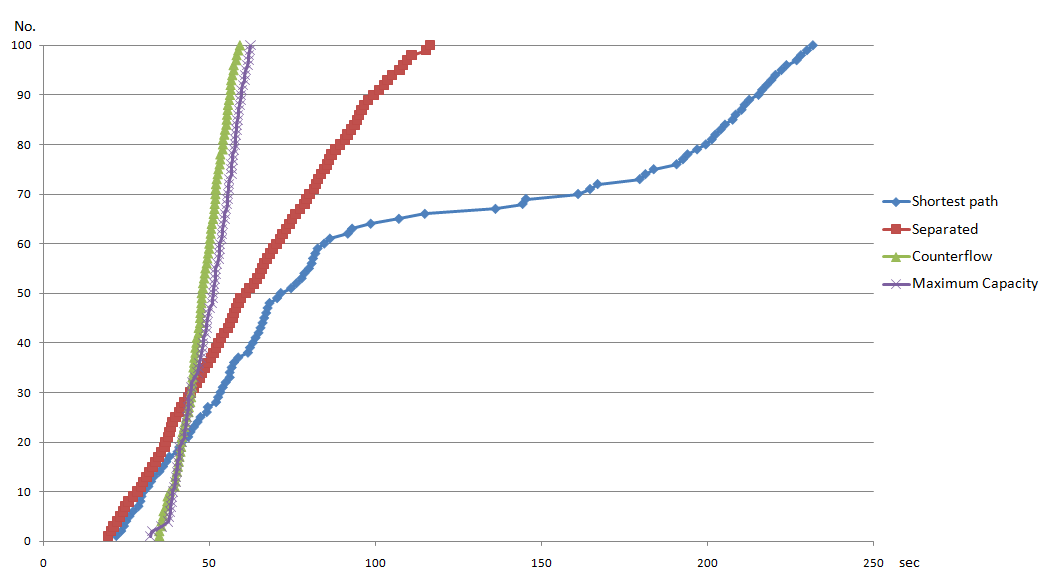}
\caption{Number of red agents that have arrived at their destination in dependence of the simulation time (evacuation graph). This plot shows that in the beginning it is faster for members of the red group to use exit A (as they are the first to arrive there), however if a red agent is not among the very first of his group then exit B is the better choice. The delay in the ``shortest path'' strategy curve exists as in this run many blue agents passed exit A before the red agents joined.  }%
\label{fig:evacred}%
\end{center}
\end{figure}

\begin{figure}[htbp]%
\begin{center}
\includegraphics[width=0.618\columnwidth]{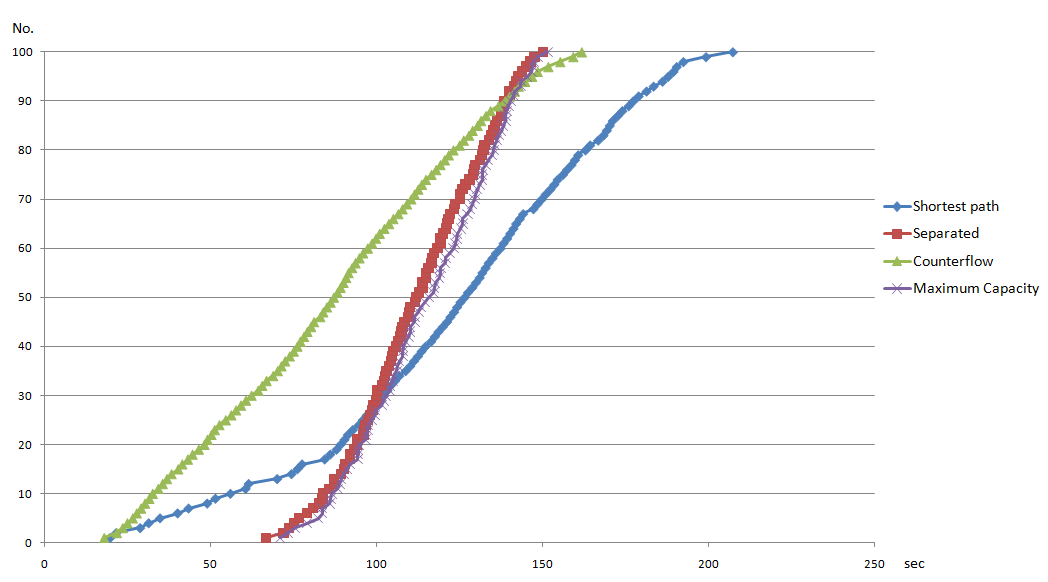}
\caption{Number of blue agents that have arrived at their destination in dependence of the simulation time (evacuation graph). From this plot it becomes clear that the counterflow strategy benefits the blue agents most.}%
\label{fig:evacblue}%
\end{center}
\end{figure}

Figures \ref{fig:evacboth} to \ref{fig:evacblue} show the number of agents that have arrived at their destination area in dependence of the simulation time. Shown are not averages, but the time dependence of those simulation runs from which the figures \ref{fig:20sec} to \ref{fig:120sec} were taken and which with their average individual egress time are closest to the average of all simulation runs.

\section{Discussion and Summary}
It has been shown by thought and simulation that situations, circumstances, and evaluation measures exist with which the quickest evacuation can be achieved when the egress procedure includes a counterflow. Both, thought as well as simulation, were carried out with simplifications. If route choice were not bound to whole groups, but could be carried out and optimized individually clearly some other result would follow. However, it appears to be not far fetched to assume that still {\em parts} of the blue group would find themselves in a situation of counterflow with {\em parts} of the red group, which would imply that the main claim of the contribution would still be correct. 

For this contribution the corridor width was chosen wide enough such that the friction stemming from counterflow between the occupants of the two groups is small. However, the results as shown in table \ref{tab:results1} are clear enough to allow for increased counterflow friction. As different models of pedestrian dynamics yield strongly different friction for identical inflow and corridor width, the case with small friction is the one were models agree most and at the same time can be assumed to agree most with a -- still to be done -- experiment.

The impact of this study for application is currently very small. Present day emergency egress signage normally is static and group independent. To implement the solution as proposed by this study, however, an emergency egress signage would have to tell ``if you are a fast walker walk this way, else the other way´´, which to date is too complicated. A possible application thus will probably have to wait until individual routing information can be calculated reliably and quickly and given via mobile devices to each occupant individually in a comprehensible way \cite{Kluepfel2010a} . 

The lessons that nevertheless can be learned from the result presented in this contribution is -- once more -- to mistrust simple truths, after all there could be other situations, where it is not so unrealistic to place an adequate emergency egress signage. Second, the fundamental condition for the results as presented is that the two groups have different (desired) walking speeds. Therefore this is another example that the phenomena of pedestrian dynamics depend much on the width of the distribution of (desired) walking speeds, as it is the case for the stability of lanes in counterflows and stripes in crossing flows. Most obviously, however, the example is a hint that with largely scattered (desired) walking speeds there are new phenomena to be considered in the search for quickest travel time solutions.

A future elaboration of the study is to do route choice not for groups, but individually. Route choice based on estimated remaining travel time has recently been in the focus of a number of studies \cite{Kretz2009a,Kirik2009,Kretz2009c,Kretz2010c,Kretz2010x,Ondrej2010,Dressler2010,Hoecker2010,Moussaid2011,Kemloh2011,Kretz2011x}. The methods proposed in these works could be used to compute a travel time based route choice individually.

\section{Supplemental Material}
The animations from which the screen shots of figures \ref{fig:20sec} to \ref{fig:120sec} were taken are available online: \url{http://www.youtube.com/watch?v=-cXH6ExUTg4}

\nocite{_PED2001,_PED2010}

\renewcommand{\bibnumfmt}[1]{#1.}

\setlength{\bibsep}{0pt}

\bibliographystyle{utphys2011b}

\bibliography{counterflow_in_evacuations}

\providecommand{\href}[2]{#2}\begingroup\begin{thebibliography}{10}

\bibitem{Simon1998}
P.~Simon and H.~Gutowitz, ``A cellular automaton model for bi-directional
  traffic'', {\em ArXiv Statistical Mechanics e-prints} (1998) 0,
  \href{http://arxiv.org/abs/9801024v1}{{\ttfamily arXiv:9801024v1
  [cond-mat]}}.

\bibitem{Muramatsu1999}
M.~Muramatsu, T.~Irie, and T.~Nagatani, ``{Jamming transition in pedestrian
  counter flow}'', \href{http://dx.doi.org/10.1016/S0378-4371(99)00018-7}{{\em
  Physica A} {\bfseries 267} (1999) 487--498}.

\bibitem{Blue2000}
V.~Blue and J.~Adler, ``{Cellular Automata Microsimulation of Bi-Directional
  Pedestrian Flows}'',
  \href{http://dx.doi.org/10.1016/S0191-2615(99)00052-1}{{\em Transportation
  Research Record, Journal of the Transportation Research Board} {\bfseries
  1678} (2000) 135--141}.

\bibitem{Algadhi2001}
S.~AlGadhi, H.~Mahmassani, and R.~Herman, ``{A Speed-Concentration Relation for
  Bi-Directional Crowd Movements with Strong Interaction}'', in Schreckenberg
  and Sharma \cite{_PED2001}, pp.~3--20.
\newblock {ISBN:3-540-42690-6}.

\bibitem{Tajima2002}
Y.~Tajima, K.~Takimoto, and T.~Nagatani, ``{Pattern formation and jamming
  transition in pedestrian counter flow}'',
  \href{http://dx.doi.org/10.1016/S0378-4371(02)00965-2}{{\em Physica A}
  {\bfseries 313} (2002) 709--723}.

\bibitem{Isobe2004}
M.~Isobe, T.~Adachi, and T.~Nagatani, ``{Experiment and simulation of
  pedestrian counter flow}'',
  \href{http://dx.doi.org/10.1016/j.physa.2004.01.043}{{\em Physica A}
  {\bfseries 336} (2004) 638--650}.

\bibitem{Kretz2006d}
T.~Kretz, A.~Gr\"unebohm, M.~Kaufman, F.~Mazur, and M.~Schreckenberg,
  ``{Experimental study of pedestrian counterflow in a corridor}'',
  \href{http://dx.doi.org/10.1088/1742-5468/2006/10/P10001}{{\em Journal of
  Statistical Mechanics: Theory and Experiment} {\bfseries 2006} (2006)
  P10001}, \href{http://arxiv.org/abs/0609691v1}{{\ttfamily arXiv:0609691v1
  [cond-mat]}}.

\bibitem{Kretz2008d}
T.~Kretz, M.~Kaufman, and M.~Schreckenberg,
  \href{http://dx.doi.org/10.1007/978-3-540-79992-4_75}{``{Counterflow
  Extension for the F.A.S.T.-Model}'',} in {\em {Cellular Automata -- 8th
  International Conference on Cellular Automata for Research and Industry, ACRI
  2008, Proceedings}}, H.~Umeo, S.~Morishita, K.~Nishinari, T.~Komatsuzaki, and
  S.~Bandini, eds., pp.~555--558.
\newblock Springer-Verlag, Berlin Heidelberg, 2008.
\newblock \href{http://arxiv.org/abs/0804.4336}{{\ttfamily arXiv:0804.4336
  [cs.MA]}}.
\newblock {ISBN:978-3-540-79991-7}.

\bibitem{Schadschneider2009}
A.~Schadschneider, W.~Klingsch, H.~Kl{\"u}pfel, T.~Kretz, C.~Rogsch, and
  A.~Seyfried,
  \href{http://dx.doi.org/10.1007/978-1-4419-7695-6_29}{``{Evacuation Dynamics:
  Empirical Results, Modeling and Applications}'',} in {\em {Encyclopedia of
  Complexity and System Science}}, R.~Meyers, ed., p.~3142.
\newblock Springer Science+Business Media, New York, June, 2009.
\newblock \href{http://arxiv.org/abs/0802.1620}{{\ttfamily arXiv:0802.1620
  [physics.soc-ph]}}.
\newblock {ISBN:978-0-387-75888-6}.

\bibitem{Schadschneider2009b}
A.~Schadschneider, H.~Kl{\"u}pfel, T.~Kretz, C.~Rogsch, and A.~Seyfried,
  \href{http://dx.doi.org/10.4018/978-1-60566-226-8.ch006}{``{Fundamentals of
  Pedestrian and Evacuation Dynamics}'',} in {\em {Multi-Agent Systems for
  Traffic and Transportation Engineering}}, A.~Bazzan and F.~Kl{\"u}gl, eds.,
  ch.~{VI}, pp.~124--154.
\newblock Information Science Reference, Hershey, PA, USA, 2009.
\newblock ISBN:978-1-60566-226-8.

\bibitem{Bell1997}
M.~Bell and Y.~Iida, {\em Transportation network analysis}.
\newblock Wiley, 1997.
\newblock {ISBN:978-0-471-96493-3}.

\bibitem{IMOMSC1238}
{IMO correspondence group}, ``{Interim Guidelines for Evacuation Analyses for
  New and Existing Passenger Ships}'', Tech. Rep. {MSC/Circ. 1238},
  International Maritime Organisation (IMO), 2007.

\bibitem{Kretz2008b}
T.~Kretz, S.~Hengst, and P.~Vortisch, ``{Pedestrian Flow at Bottlenecks --
  Validation and Calibration of VISSIM's Social Force Model of Pedestrian
  Traffic and its Empirical Foundations}'', in {\em {International Symposium of
  Transport Simulation 2008 (ISTS08)}}, M.~Sarvi, ed., p.~electronic
  publication.
\newblock Monash University, Gold Coast, Australia, 2008.
\newblock \href{http://arxiv.org/abs/0805.1788}{{\ttfamily arXiv:0805.1788
  [cs.MA]}}.

\bibitem{Fellendorf2010}
M.~Fellendorf and P.~Vortisch, ``Microscopic Traffc Flow Simulator VISSIM'',
  \href{http://dx.doi.org/10.1007/978-1-4419-6142-6_2}{{\em Fundamentals of
  Traffic Simulation} (2010) 63--94}. \url{http://tinyurl.com/3quwaqz}.

\bibitem{VISSIM2010}
PTV, {\em {VISSIM 5.30 User Manual}}.
\newblock PTV Planung Transport Verkehr AG, Stumpfstra{\ss}e 1, D-76131
  Karlsruhe, 2010.
\newblock \url{http://www.vissim.de/}.
\newblock The simulations have been carried out with version 5.30-05. A free
  trial version with sufficient functionality with regard to the content of
  this paper can be downloaded from the given webpage.

\bibitem{Johansson2007}
A.~Johansson, D.~Helbing, and P.~Shukla, ``{Specification of the Social Force
  Pedestrian Model by Evolutionary Adjustment to Video Tracking Data}'',
  \href{http://dx.doi.org/10.1142/S0219525907001355}{{\em Advances in Complex
  Systems} {\bfseries 10} no.~4, (2007) 271--288},
  \href{http://arxiv.org/abs/0810.4587}{{\ttfamily arXiv:0810.4587
  [physics.soc-ph]}}.

\bibitem{Helbing2009}
D.~Helbing and A.~Johansson,
  \href{http://dx.doi.org/10.1007/978-1-4419-7695-6_37}{``{Pedestrian, Crowd
  and Evacuation Dynamics}'',} in {\em {Encyclopedia of Complexity and System
  Science}}, R.~Meyers, ed., vol.~16, p.~6476.
\newblock Springer Science+Business Media, New York, 2009.
\newblock \url{http://tinyurl.com/3lul642}.
\newblock {ISBN:978-0-387-75888-6}.

\bibitem{Kreps1992}
D.~Kreps, ``Game theory and economic modelling'', {\em OUP Catalogue} (1992) 0.

\bibitem{Camerer2003}
C.~Camerer and R.~S. Foundation, {\em Behavioral game theory: Experiments in
  strategic interaction}, vol.~9.
\newblock Princeton University Press Princeton, NJ, 2003.

\bibitem{Osborne2004}
M.~Osborne, {\em An introduction to game theory}, vol.~3.
\newblock Oxford University Press New York, NY, 2004.

\bibitem{Kretz2011d}
T.~Kretz, ``{A Round-Robin Tournament of the Iterated Prisoner's Dilemma with
  Complete Memory-Size-Three Strategies}'', {\em {Complex Systems}} {\bfseries
  19} no.~4, (2011) 363--389, \href{http://arxiv.org/abs/1101.0340}{{\ttfamily
  arXiv:1101.0340 [cs.GT]}}.
  \url{http://www.complex-systems.com/pdf/19-4-4.pdf}.

\bibitem{Kluepfel2010a}
H.~Kl{\"u}pfel, A.~Seyfried, S.~Holl, M.~Boltes, M.~Chraibi, U.~Kemloh,
  A.~Portz, J.~Liddle, T.~Rupprecht, A.~Winkens, W.~Klingsch, C.~Eilhardt,
  S.~Nowak, A.~Schadschneider, T.~Kretz, and M.~Krabbe, ``{HERMES -- Evacuation
  Assistant for Arenas}'', in {\em Future Security 2010}.
\newblock Fraunhofer VVS, 2010.
\newblock eprint.

\bibitem{Kretz2009a}
T.~Kretz, ``{Pedestrian Traffic: on the Quickest Path}'',
  \href{http://dx.doi.org/10.1088/1742-5468/2009/03/P03012}{{\em Journal of
  Statistical Mechanics: Theory and Experiment} {\bfseries 2009} (2009)
  P03012}, \href{http://arxiv.org/abs/0901.0170}{{\ttfamily arXiv:0901.0170
  [physics.soc-ph]}}.

\bibitem{Kirik2009}
E.~Kirik, T.~Yurgel'yan, and D.~Krouglov, ``The shortest time and/or the
  shortest path strategies in a CA FF pedestrian dynamics model'', {\em Journal
  of Siberian Federal University. Mathematics \& Physics} {\bfseries 2} no.~3,
  (2009) 271--278, \href{http://arxiv.org/abs/0906.4265}{{\ttfamily
  arXiv:0906.4265 [math-ph]}}.

\bibitem{Kretz2009c}
T.~Kretz, ``{The use of dynamic distance potential fields for pedestrian flow
  around corners}'', in {\em First International Conference on Evacuation
  Modeling and Management}.
\newblock TU Delft, 2009.
\newblock \href{http://arxiv.org/abs/0804.4336}{{\ttfamily arXiv:0804.4336
  [cs.MA]}}.

\bibitem{Kretz2010c}
T.~Kretz, ``{Applications of the Dynamic Distance Potential Field Method}'', in
  {\em {Traffic and Granular Flow '09}}, {Dai, S. et al.}, ed., p.~0.
\newblock Springer Berlin Heidelberg, 2009.
\newblock \href{http://arxiv.org/abs/0911.3723}{{\ttfamily arXiv:0911.3723
  [cs.MA]}}.
\newblock in press.

\bibitem{Kretz2010x}
T.~Kretz, \href{http://dx.doi.org/10.1007/978-3-642-15979-4_51}{``{The Dynamic
  Distance Potential Field in a Situation with Asymmetric Bottleneck
  Capacities}'',} in {\em Cellular Automata -- 9th International Conference on
  Cellular Automata for Research and Industry, ACRI 2010}, S.~Bandini,
  S.~Manzoni, H.~Umeo, and G.~Vizzari, eds., vol.~6350 of {\em Lecture Notes in
  Computer Science}, pp.~480--488.
\newblock Springer, Heidelberg, 2010.
\newblock {ISBN:978-3-642-15978-7}.

\bibitem{Ondrej2010}
J.~Ond{\v{r}}ej, J.~Pettr{\'e}, A.~Olivier, and S.~Donikian, ``A
  synthetic-vision based steering approach for crowd simulation'',
  \href{http://dx.doi.org/10.1145/1778765.1778860}{{\em ACM Transactions on
  Graphics (TOG)} {\bfseries 29} no.~4, (2010) 1--9}.

\bibitem{Dressler2010}
D.~Dressler, M.~Gro{\ss}, J.~Kappmeier, T.~Kelter, J.~Kulbatzki, D.~Pl{\"u}mpe,
  G.~Schlechter, M.~Schmidt, M.~Skutella, and S.~Temme, ``{On the use of
  network flow techniques for assigning evacuees to exits}'',
  \href{http://dx.doi.org/10.1016/j.proeng.2010.07.019}{{\em Procedia
  Engineering} {\bfseries 3} (2010) 205--215}.

\bibitem{Hoecker2010}
M.~H{\"o}cker, V.~Berkhahn, A.~Kneidl, A.~Borrmann, and W.~Klein, ``Graph-based
  approaches for simulating pedestrian dynamics in building models'', in {\em
  8th European Conference on Product \& Process Modelling (ECPPM), University
  College Cork, Cork, Ireland}.
\newblock 2010.
\newblock \url{http://tinyurl.com/3gn8qnk}.

\bibitem{Moussaid2011}
M.~Moussa{\"i}d, D.~Helbing, and G.~Theraulaz, ``How simple rules determine
  pedestrian behavior and crowd disasters'',
  \href{http://dx.doi.org/10.1073/pnas.1016507108}{{\em Proceedings of the
  National Academy of Sciences} {\bfseries 108} no.~17, (2011) 6884}.

\bibitem{Kemloh2011}
A.~{Kemloh Wagoum}, A.~Seyfried, and S.~Holl, ``Modelling dynamic route choice
  of pedestrians to assess the criticality of building evacuation'',
  \href{http://arxiv.org/abs/1103.4080}{{\ttfamily arXiv:1103.4080 [cs.OH]}}.
  submitted.

\bibitem{Kretz2011x}
T.~Kretz, A.~Gro{\ss}e, S.~Hengst, L.~Kautzsch, A.~Pohlmann, and P.~Vortisch,
  ``{Quickest Paths in Simulations of Pedestrians}'', {\em Advances in Complex
  Systems} (2011) 0, \href{http://arxiv.org/abs/1107.2004}{{\ttfamily
  arXiv:1107.2004 [physics.soc-ph]}}. in press.

\bibitem{_PED2001}
M.~Schreckenberg and S.~Sharma, eds., {\em {Pedestrian and Evacuation
  Dynamics}}.
\newblock Springer-Verlag Berlin Heidelberg, Duisburg, 2002.
\newblock {ISBN:3-540-42690-6}.

\bibitem{_PED2010}
{Peacock, R.D., Kuligowski, E.D., and Averill, J.}, ed.,
  \href{http://dx.doi.org/10.1007/978-1-4419-9724-1}{{\em Pedestrian and
  Evacuation Dynamics}}.
\newblock Springer, 2011.
\newblock {ISBN:978-1-4419-9724-1}.

\end{thebibliography}\endgroup

\end{document}